\documentclass[sigconf]{acmart}

\AtBeginDocument{%
  }
\usepackage{enumitem}
\usepackage{multirow}
\usepackage{stfloats}
\usepackage[section]{placeins}
\usepackage{float}

\copyrightyear{2025}
\acmYear{2025}
\setcopyright{cc}
\setcctype{by-nc}
\acmConference[CHI '25]{CHI Conference on Human Factors in Computing Systems}{April 26-May 1, 2025}{Yokohama, Japan}
\acmBooktitle{CHI Conference on Human Factors in Computing Systems (CHI '25), April 26-May 1, 2025, Yokohama, Japan}\acmDOI{10.1145/3706598.3714100}
\acmISBN{979-8-4007-1394-1/25/04}

\newcommand\anedit[1]{#1}

\newcommand\anold[1]{#1}

\begin{document}

\title[Knowledge Workers' Perspectives on AI Training for Responsible AI Use]{Knowledge Workers' Perspectives on AI Training for Responsible AI Use}




\author{Angie Zhang}
\affiliation{
  \institution{School of Information \\ The University of Texas at Austin}
   \city{Austin}
   \state{TX}
  \country{United States}}
\email{angie.zhang@austin.utexas.edu}
\orcid{}

\author{Min Kyung Lee}
\affiliation{
  \institution{School of Information \\ The University of Texas at Austin}
   \city{Austin}
   \state{TX}
  \country{United States}
}
\email{minkyung.lee@austin.utexas.edu}

\renewcommand{\shortauthors}{Zhang \& Lee}

\begin{abstract}
AI expansion has accelerated workplace adoption of new technologies. Yet, it is unclear whether and how \anold{knowledge} workers are supported and trained to safely use AI. Inadequate training may lead to unrealized benefits if workers abandon tools, or perpetuate biases if workers misinterpret AI-based outcomes. In a workshop with 39 workers from 26 countries specializing in human resources, labor law, standards creation, and worker training, we explored questions and ideas they had about safely adopting AI. We held 17 follow-up interviews to further investigate what skills and training \anold{knowledge} workers need to achieve safe and effective AI in practice. \anold{We synthesize nine training topics participants surfaced for knowledge workers related to challenges around \anold{understanding what AI is}, \anold{misinterpreting outcomes}, \anold{exacerbating biases},} and worker rights. We reflect how these training topics might be addressed under different contexts, imagine HCI research prototypes as potential training tools, and consider ways to ensure training does not perpetuate harmful values.

\end{abstract}



\maketitle

\section{Introduction}
Organizations are racing to adopt AI, \anold{especially generative AI tools,} for their promise to transform industries \cite{deloitte_2023, lane2023impact, mckinsey_2023, brynjolfsson2023generative}---e.g., automating repetitive tasks and freeing workers for more "innovative" or critical tasks. Such pursuits must also consider potential consequences of hasty AI deployment such as workers over-relying on AI and perpetuating incorrect or biased outcomes \cite{bansal2021does}, the deskilling of workers resulting in their replacement by machines \cite{brynjolfsson2024navigating}, and workers abandoning beneficial tools due to insufficient understanding about how to use them \cite{allen2022algorithm}. To ensure effective and safe AI adoption, reports and articles urge for AI-capable workforces by way of training to \textit{upskill} employees to use AI effectively and \textit{reskill} employees whose job functions may be replaced by AI \cite{deloitte_2023, mckinsey_2023, shiohira2021understanding, brynjolfsson2024navigating}. 

\anedit{Researchers and organizations have released early reports that synthesize a wide range of expected or desired skills for an AI-capable workforce \cite{lane2023impact, shiohira2021understanding, morandini2023impact, deloitte_2023, mckinsey_2023}. \anedit{These reports underscore the importance of technical skills such as advanced data analytics, programming and modeling, experience with specific AI tools, and different literacies to interpret AI outcomes.} Reports also emphasize that workers need strong \textit{non-technical} skills in communication, critical thinking, and self-management. Lastly, researchers and organizations highlight that workers should receive training on the social and ethical implications of AI.}

Notably though, \textbf{many reports and articles cataloging what skills and topics workers need often center the perspectives or desires of \textit{\anedit{leadership} or \anedit{senior management}}} \cite{lane2023impact, mckinsey_2023, russell2023competencies, shiohira2021understanding, squicciarini2021demand, alekseeva2021demand} rather than workers \anedit{in non-executive roles} who are expected to use AI in their work\anedit{---including individual contributors and lower-level management roles}. 
Additionally, while related literature frequently calls for worker training to build competencies of an AI-capable workforce, it is still unclear \textit{what} training workers are currently receiving---e.g., the topics, training formats, and efficacy of training. Researchers have called attention to the ongoing need for understanding \textit{how} upskilling or reskilling should occur. For example, \citet{woodruff2024knowledge}'s recent findings revealed knowledge workers are concerned about being deskilled by AI, and the authors encourage the HCI community to investigate what training workers require in this rapidly evolving technological era. 

To address these open questions \anold{around training and what workers need}, we conducted a workshop (n=39 participants) and semi-structured interviews (n=17 of the original 39) to learn workers' experiences, concerns, and ideas for training on AI. 
\anold{AI experiences and training topics can depend on factors such as workplace and role, AI application type, and training phase. We focus on experiences of information and knowledge workers\footnote{Hereafter, we refer to information and knowledge workers collectively as "knowledge workers."} and the AI tools they are using, including "general-purpose AI systems"---AI that can solve "a wide array of tasks without being specifically designed for them" \cite{triguero2024general}. Given many general-purpose AIs are being designed to perform tasks replicating or assisting the roles of knowledge workers and being adopted by workplaces, understanding this group of workers and technology is very topical. Furthermore, we aimed to address overall training, rather than one specific training phase, to address ongoing calls for general workplace AI readiness.} 

%
\anold{Through our study, we} sought to understand the following research questions:

\begin{enumerate}
    \item[RQ1:] How do \anold{knowledge} workers currently receive training and support for using AI at work? 
    \item[RQ2:] What are \anold{knowledge} workers' perspectives about training topics and skills they need to use AI effectively and safely at work?
\end{enumerate}

Our paper contributes to emerging work in the HCI domain about supporting responsible AI education for \anold{knowledge} workers. We synthesize the experiences, challenges, and ideas of a multinational group of workers to \anold{identify four challenges participants feel knowledge workers need AI education or training for:

\begin{itemize}
    \item \anold{\textbf{Challenge 1:} } \anold{Some workers do not know what AI is or how it could help them with work tasks, which can risk \anedit{them missing out on career opportunities} in a rapidly evolving workforce \anedit{that requires new skills}.}
    \item \anold{\textbf{Challenge 2:} } \anold{Workers risk over-relying on AI and blindly trusting its outputs which can lead to harmful outcomes and loss of critical thinking skills.}
    \item \anold{\textbf{Challenge 3:} } \anold{Workers may not be sufficiently aware about how AI can exacerbate DEI risks and perpetuate biases.}
    \item \anold{\textbf{Challenge 4:} } \anold{Workers are concerned about how AI may infringe on worker rights and data privacy.}
\end{itemize}

For each challenge, we explain related training topics they identified to support} effective and responsible AI adoption\anold{, resulting in a total of nine training topics.} \anold{Our findings indicate the need to support knowledge workers by grounding their training in a fundamental understanding of what AI is \textbf{(T1-T2)}, reframing AI products as tools and not solutions \textbf{(T3-T4)}, emphasizing AI risks and benefits on DEI as a priority and not an afterthought \textbf{(T5-T6)}, and championing worker empowerment \textbf{(T7-T9)}.}
Based on our findings, we reflect how these training \anold{topics} can be addressed in three ways: 1) in consideration of different contexts that may impact how training \anold{topics} are supported, 2) by offering one idea for how ongoing and future HCI research and prototypes could support training \anold{topics}, and 3) by being mindful of how values can become embedded in technology through training. \anold{Though we present these as training topics and implications for knowledge workers, they may be useful as guidance for organizations of other worker types facing AI deployments.}


\section{Understanding How AI Training or Education for Workers Is Currently Supported in the Workplace}\label{related_work}
To understand how workers are currently supported in training or education on AI in the workplace, we turn to two related research domains: 1) AI literacy and 2) responsible AI development, design, and deployment. The former offers a view into how AI education and related skills/topics have been formulated and disseminated. The latter offers a glimpse into resources that researchers have created to support responsible AI efforts and how workers may be learning on the job.

\subsection{AI Literacy for Workers}\label{related_work_literacy}
Researchers often define AI literacy through a lens of the competencies users need so they can understand and use AI in their lives. For example, \citet{long2020ai} define AI literacy as "a set of competencies that enables individuals to critically evaluate AI technologies; communicate and collaborate effectively with AI; and use AI as a tool online, at home, and in the workplace." Other reviews on AI literacy have surfaced that researchers often study how to support four types of user needs: 1) knowing and understanding basic functions, 2) applying this knowledge, 3) critically examining or developing AI, and 4) understanding AI ethics \cite{ng2021ai, dominguez2023responsible}.

To help individuals develop AI literacy, researchers have explored formal and informal methods of teaching AI. Formal methods have included creating curriculum to introduce general fundamentals of AI and development of modules to teach related technical skills (e.g., programming languages, data analytics) \cite{lee2021developing, how2019educing}. Informal methods include creating AI-driven artefacts or interactions for learners to engage and explore \cite{long2021role} or pedagogical activities that encourage exploration of AI concepts or AI-powered tools, such as digital story writing \cite{ng2022using}.
For example, \citet{lee2021developing} created an AI literacy curriculum for middle school students that teaches key AI concepts (e.g., machine learning, logic systems), \anedit{incorporates AI ethics and career development activities}, and gives hands-on practice such as building projects using Scratch programming. Exploring informal spaces and methods of learning, \citet{long2021role} developed three interactive, embodied museum exhibits. They evaluated how families learned about AI using these, for example, finding learners reported a \anedit{gain} in knowledge about "types of AI and the differences between them" after using an interface to manipulate gestures of an AI dance partner. 

Broadly, this area of research has focused on supporting the AI literacy of two populations: the general (non-expert) public to support an understanding of AI (e.g., \cite{long2020ai, long2021co, laupichler2023delphi, markauskaite2022rethinking}); and teachers and students in K-12 or at the university-level (e.g., AI for Education, \cite{tan2024more, mathew2024needs, long2021co, kandlhofer2016artificial, lee2021developing, ng2022using}). \anedit{Research} has been conducted to examine how other user groups, including different domains of workers, understand AI or literacy support they need \cite{charow2021artificial, pinski2024ai, cetindamar2022explicating}. For example, \citet{cetindamar2022explicating} conducted a bibliometrics review of 270 articles related to AI literacy and workers to outline that capabilities necessary for "work-related AI literacy" primarily relate to general use of AI in the workplace and related skills for effective AI deployment. However, research on \textit{how} to support AI literacy or AI training of other groups---including workers---is largely overlooked. This is a concern given the volume of people who are part of the workforce: the Organization for Economic Co-operation and Development (OECD) reported a labor force participation rate of 73.8\% across its 38 member states' populations in 2023 \cite{OECD_2024}. 

One exception that \textit{does} specifically support workers is \citet{kaspersen2024primary}\anedit{---the authors} draw from AI literacy tools for students to develop a tool that teaches workers of a trade union about AI. Another adjacent effort to AI literacy or training for workers is \citet{kotturi2024deconstructing}'s unique workshops \anedit{that} onboarded tech entrepreneurs to use generative AI to further their businesses. Both are excellent points of reference in the efforts towards AI literacy and training for workers.

\subsection{Supporting Responsible AI in Workplaces: Design, Development, and Deployment}\label{related_work_responsible}
Towards the responsible use of AI in the workplace, HCI and CSCW researchers often investigate how to support workers in AI design, development, and deployment. One thread of research has focused on creating technical tools or toolkits to help developers benchmark and visualize AI performance as related to fairness and ethical concerns (e.g., \cite{saleiro2018aequitas, bird2020fairlearn, bellamy2019ai}). These tools are largely aimed at assisting technical practitioners such as engineers, developers, and data scientists as they require technical expertise to implement and understand such metrics. 
%
Researchers and organizations have also created resources and toolkits to support technical and non-technical practitioners in designing AI responsibly, especially as it relates to facilitating critical reflection about ethical implications of AI deployment \cite{kawakami2024situate, shen2021value, crisan2022interactive, IDEO}.
For example, \citet{kawakami2024situate} proposed a guidebook of questions for public sector employees to use when deliberating and reflecting on AI adoption. \citet{shen2021value} created educational cards to help college students learn about the ethical impacts of machine learning (ML). Different tools have also been created to facilitate critical reflection about responsible AI amongst the practitioners developing or engineering these systems \cite{bender2018data, gebru2021datasheets, madaio2020co}. 

\anold{Related to helping workers learn to use AI responsibly, one area has studied how to support workers' proper interpretations of AI recommendations \cite{cai2019hello, mozannar2022teaching, kawakami2023training}. Researchers focus on how interfaces, training, and feedback can help workers calibrate their mental models of AI decision-making to improve outcomes of human-AI collaboration. For example, \citet{cai2019hello} investigated what medical experts need to know about an AI during the onboarding phase---"when users are first being introduced to an AI system, learning its capabilities, and determining how they will partner with it in practice"---to use it appropriately. \citet{cai2019hello} summarized five aspects experts need to calibrate their mental models of AI tools---its capabilities, data it uses and how, severity of its judgment, what metrics it is optimized for, and what factors were considered for adopting it. In line with supporting users' mental models, \citet{mozannar2022teaching} suggested doing so by helping participants accurately interpret an AI's strengths (what it predicts correctly) and weaknesses (what it gets wrong) to improve participants' performance on classification tasks.} \anold{\citet{kawakami2023training} similarly investigated how training---in this case, repeated practice---with an assistive AI decision-making tool can help users learn when to critically disagree with AI predictions.} \anold{These studies contribute to our understanding around how to onboard workers in using AI assistants responsibly, although they primarily focus on the context of workers using task-specific decision-making tools. AI work tools for other (non-predictive) tasks or purposes may result in different implications for training.}

Another \anold{related} area \anold{has focused} on helping workers learn informally about AI \anold{related to} workplace surveillance and algorithmic management. This has often occurred in the context of platform gig workers. For example, \citet{do2024designing} co-designed new technologies that advance sousveillance---"the act of subordinates monitoring people in power"---enabling workers to simultaneously design and learn about workplace surveillance. Similar approaches that allow workers to increase their awareness about AI for management encompass worker-centered auditing efforts such as \citet{calacci2022bargaining} partnering with a non-profit worker organization to reverse engineer Shipt's delivery platform payment algorithm and assess if it was negatively impacting workers' earnings. 

Finally, aside from resources and design activities listed above, researchers are exploring how workers learn responsible AI \textit{on the job} \cite{varanasi2023currently, madaio2024learning, berman2024scoping}. This has surfaced limitations such as few workers receiving formal training and most workers depending on informal conversations with others and self-learning methods, e.g., independently seeking out online courses, research papers, or relevant monographs \cite{madaio2024learning}. \citet{varanasi2023currently} spoke to participants who shared similar experiences and emphasized that responsible AI initiatives fall on a small subset to carry out. Participants also described challenges \anedit{holding} meaningful deliberation \anedit{with} colleagues when trying to implement responsible AI values---suggesting that reflective or deliberative resources such as those mentioned earlier may fall short or be stilted by situational and organizational factors. An important distinction here is that these studies have focused on the informal, self-learning methods of those developing or designing responsible AI as opposed to non-AI practitioners learning to \textit{use} AI responsibly, which is the focus of our study.



\section{Method}\label{method}

\subsection{Background}
In \anold{July} 2024, the primary author facilitated a workshop as part of a learning series hosted by the International Training Centre of the International Labour Organization (ITCILO) on \textit{AI in the Workplace}.
The International Labour Organization (ILO) is a United Nations agency that "brings together governments, employers and workers of 187 Member States, to set labour standards, develop policies and devise programmes promoting decent work for all women and men."\footnote{https://www.ilo.org/about-ilo} The ITCILO is the training arm of the ILO. Its mission is "to provide people with access to digitally-enhanced capacity development services to successfully manage their future of work transitions."\footnote{https://www.itcilo.org/training} The organization has a history of creating and providing learning and training to participants from different countries and professions. 

The workshop served as an interactive learning activity that introduced participants to types of workplace AI to identify gaps in their understanding of AI in the workplace and what they need/want to know to use it responsibly. 
During the workshop, the primary author observed how participants often discussed the need for training so workers can safely use and interpret AI outputs, as well as the importance of workplaces providing transparency about why AI is being adopted.
Inspired by their insights, the primary author reached out to all participants about follow-up interviews to expand on these topics with their perspectives and firsthand experiences. 

\subsection{Workshop}
39 individuals participated \anold{in the workshop}, representing 26 unique countries from 5 continents. The workshop began by introducing workers to a scenario-based activity. \anedit{To familiarize them and allow for clarifying questions,} participants were guided through an example with pre-populated sticky notes. 
Next, participants were evenly divided into three \anold{virtual} breakout groups with one facilitator each---the primary author and the two organizers \anold{of the learning series}. \anold{In their breakout groups,} participants worked on a scenario-based activity \anold{that described} an AI tool being used by a workplace \anold{by} brainstorming ideas and questions about the scenario and tool, guided by prompts. 

The AI depicted were \anedit{based on} real, workplace AI \anold{products}: a tracking and scheduling tool, warehouse scanners,  and an assistive AI for performance evaluation. Each use case was mapped to a worker rights issue---e.g., the assistive AI for performance evaluation was mapped to "Bias/Discrimination"---because one objective of the learning series was to help participants understand risks AI can pose to workers' rights.\footnote{The primary author sought feedback from the organizers iteratively to finalize the scenarios.} (See Table \ref{table_workshop_scenarios} for the full breakdown.) The prompts included to help workers think critically were: \textit{What concerns or questions do you have about how this AI works?}, \textit{What would you ask vendors selling this AI and/or workplaces that procure AI?}, and \textit{What are ideas about how to improve the design of this AI?}


We acknowledge that the scenarios and prompts could have influenced the responses and ideas of participants about training \anold{topics}. However, we believed that introducing frictions of these potential problems and concerns was important to encourage participants to think critically. Literature \anold{has shown} that \anold{reflecting over technology} ethics and biases is not natural for many, not often mandated by workplaces, and frequently falls on the shoulders of the few \cite{madaio2024learning, varanasi2023currently, rakova2021responsible}. Thus, intentionally \anold{probing} participants on issues and topics such as diversity, equity, and inclusion (DEI) is important. Technological impact on these may \anold{otherwise} be overlooked as benign.

%

The workshop lasted an hour: participants spent the first half in breakout rooms and the second half as a full group \anold{to debrief} ideas and discussions. It was facilitated over Zoom and through FigJam\footnote{https://www.figma.com/figjam/}, an online collaborative whiteboard. Participants created sticky notes on FigJam \anold{for} written prompts, \anold{assisted by facilitators}. 
Within a week of the workshop's conclusion, the primary author shared documentation with participants \anold{summarizing} each group's ideas and discussions as a learning outcome and resource for participants' future reference. 

\subsection{Follow-Up Interviews}
\begin{table*}[h]
\begin{tabular}{ccccc}
\textbf{ID} & \textbf{Gender} & \textbf{Workplace Base} & \textbf{Sector}  & \textbf{Role Expertise}            \\\hline
P1                           & Female          & Africa                  & Private Industry  & \begin{tabular}[c]{@{}c@{}}HR,\\ Standards Setting\end{tabular}               \\\hline
P2                         & Female          & Africa                  & Government        & Labor Law                          \\\hline
P3                  & Female          & Europe                  & Non-Profit        & Standards Setting                  \\\hline
P4               & Female          & Latin America           & Private Industry  & \begin{tabular}[c]{@{}c@{}}HR,\\ Worker Training\end{tabular}                 \\\hline
P5               & Female          & Europe                  & IGO & Labor Law                          \\\hline
P6                & Male            & Latin America           & Government        & Labor Law                          \\\hline
P7          & Female          & South America           & Private Industry  & Labor Law                          \\\hline
P8                     & Male            & Asia                    & Private Industry  & Labor Law                          \\\hline
P9          & Female          & Europe                  & IGO & \begin{tabular}[c]{@{}c@{}}Standards Setting\\ Worker Training\end{tabular} \\\hline
P10                 & Male            & South America           & IGO & \begin{tabular}[c]{@{}c@{}}Standards Setting\\ Worker Training\end{tabular} \\\hline
P11                 & Female          & Europe                  & IGO & HR                                 \\\hline
P12              & Male            & Asia                    & Private Industry  & Worker Training                    \\\hline
P13         & Male            & Africa                  & Government        & Worker Training                    \\\hline
P14        & Female          & Europe                  & IGO & Standards Setting                  \\\hline
P15                  & Male            & Africa                  & Trade Union       & Labor Law                          \\\hline
P16         & Male            & Europe                  & Government        & Standards Setting                  \\\hline
P17         & Male            & Asia                    & Trade Union       & Worker Training                   
\end{tabular}
\captionsetup{width=.8\linewidth}
\caption{Interview Participant Table. \\ \normalfont Nationality and ethnicity are reported separately in the paper to \\protect anonymity. \\IGO = Intergovernmental Organization.}
\label{table_participant_table}
\end{table*}

The primary author reached out to workshop participants for their interest and availability in a 1-hr follow-up interview. We held interviews with 17 \anold{participants who responded}. Participants self-identified ethnicity and nationality. Of the 11 \anold{ethnicities reported}, two answered for each of the following---Black, White British, Latin American, and Asian. Remaining categories had one response each---African, Hindu, South Sulawesi. Of the 13 \anold{nationalities reported}, two answered as Japanese; two as British; and the rest (one of each) include Dutch, Indian, Indonesian, Mexican, Namibian, Nigerian, South Korean, Ugandan, and Zimbabwean. See more participant details in Table \ref{table_participant_table}.

Participants represent four areas of expertise, with some representing multiple areas. These include:
\begin{itemize}
    \item Human Resources (HR) = Staffing, recruitment, creating worker contracts, remuneration
    \item Standards Setting = Creating policies; reviewing or revising standards, policies, and/or guidelines
    \item Labor Law = Consulting on labor disputes, worker-workplace mediation, drafting text to inform policies on workplace labor standards
    \item Worker Training = Training workers on topics like DEI and workplace standards, vocational training, 1-1 coaching
\end{itemize}


During interviews, participants were asked about their background and role, and if relevant, any clarifications about comments they made during the workshop. Then the interview focused on 1) experiences using AI, 2) existing workplace practices about training on AI, 3) ideas about worker training \anold{topics} for AI and how to conduct training, and 4) perspectives about transparency workers need on AI used at work. Participants were offered a \$50 USD gift card or the equivalent in their preferred currency.
\anold{Participants' responses} reflect their personal views and are not representative of the views of their workplaces. 

%



\subsection{Analysis}
The workshop and interviews were conducted and recorded using Zoom and transcribed on Otter.ai. \anold{We followed a thematic analysis approach to guide us} \cite{patton2014qualitative,braun2006using}. \anold{We began by reviewing our data---}workshop and interview transcripts and notes\anold{---}in accordance with their respective protocols (i.e., workshop prompts and interview question areas). \anold{During this process, we began to get ideas of potential codes.} \anold{Next, we used open coding to code transcripts at the sentence or paragraph level. This allowed us to iteratively create and modify codes on segments of data relevant to our research questions while still permitting us to explore our data for additional insights and patterns.} The primary author reviewed and sorted resulting codes to surface emerging themes, and the research team met collectively to \anold{iteratively} discuss the \anold{codes and} themes. This analysis resulted \anold{in the findings reported below: challenges using AI for work and related training topics participants desire (Section \ref{findings_training}), and situational contexts that influence pursuit and outcomes of workplace training (Section \ref{findings_context}).} 


The workshop and follow-up interviews \anold{were} conducted \anold{between July to September} 2024. The results of the workshop are included with the permission of the ITCILO organizers. The interviews were an activity separate of the ITCILO's involvement, which was also explained to participants. The findings presented \anold{here} of the workshop and interviews do not represent the views of the ITCILO. 

\subsection{Research Team \& Positionality}
\anold{Our research team is based in the U.S. and} comprises individuals with academic backgrounds in human-computer interaction, design, and information science. \anold{Like our participants, we identify as knowledge workers grappling with the influx of AI for work without formal training. We are especially concerned about how to support responsible adoption of AI in practical ways. 
This motivated our curiosity to study workers' perspectives, especially from a non-U.S. perspective. We also note that our methodological approach may limit the generalization of our findings, however, we felt a qualitative approach was most appropriate to understand and contextualize the rich experiences of multinational workers interacting with AI and AI training at work.}

\anold{Additionally, we would be remiss if we did not acknowledge our privilege as researchers} in a developed country whereas a number of our participants come from developing countries. \anold{There was also} a language differential \anold{due to interviews} being conducted in English, a secondary language for some participants. This was something we treated with care, such as intentionally spending more time on questions where participants exhibited the most comfort conveying their thoughts. \anold{While analyzing the data, we found ourselves reflecting over these imbalances often, the output of which especially influenced parts of Section \ref{findings_context} where we share contexts that influence training uptake success and Section \ref{discussion_values} where we reflect on the importance of not over-privileging certain perspectives---e.g., native American English speakers, commercial enterprises.}




\section{Findings: Challenges Workers Face \& Training They Require to Support Responsible AI Use at Work}\label{findings_training}
\anold{We organize our findings into \anedit{two} sections. Here,} \anold{we explain four sets of challenges that participants described facing with workplace AI and the corresponding training topics they desire for supporting responsible workplace AI use}. 
\anold{Then in Section \ref{findings_context}, we discuss specific organizational and regional contexts that participants shared are factors that will impact the success of worker training towards effective and responsible AI adoption.}



\begin{table*}[]
\begin{tabular}{|c|c|l|l|}
\hline
\textbf{\#}         & \multicolumn{1}{c|}{\textbf{\anold{Challenge Around AI Tools for Work}}}                                                                                                                                        & \textbf{\#} & \textbf{Training \anold{Topic}}                                                                                                                                                                       \\ \hline
\multirow{3}{*}{\anold{C1}} & \multirow{2}{*}{\begin{tabular}[c]{@{}c@{}}\anold{Some workers do not know what AI is} \\ \anold{or how it could help them with work tasks, which} \\ can risk \anedit{them missing out on career opportunities} \\ \anold{in a rapidly evolving workforce \anedit{that requires new skills}.}\end{tabular}} & T1          & \begin{tabular}[c]{@{}l@{}}Establish foundational understanding\\ about what AI is and what it can do \anold{to}\\ \anold{help workers.}\end{tabular}                                                         \\ \cline{3-4} 
                    &                                                                                                                                                                                                                      & T2          & \begin{tabular}[c]{@{}l@{}}Expose workers to an array of emerging\\ AI tools \anold{to learn about new tools and the skills} \\ \anold{necessary to apply them to their own work.}\end{tabular}                     \\ \hline
\multirow{3}{*}{\anold{C2}} & \multirow{2}{*}{\begin{tabular}[c]{@{}c@{}}\anold{Workers risk over-relying on AI} \\ \anold{and blindly trusting its outputs,} \\ \anold{which can lead to harmful outcomes} \\ \anold{and loss of critical thinking skills.}\end{tabular}}         & T3          & \begin{tabular}[c]{@{}l@{}}Refresh different literacy types and\\ critical thinking skills \anold{to orient AI as a tool}\\ \anold{to assist, not a solution to rely on.}\end{tabular}                        \\ \cline{3-4} 
                    &                                                                                                                                                                                                                      & T4          & \begin{tabular}[c]{@{}l@{}}Guide workers through \anold{using their own}\\ \anold{knowledge and expertise to} critically examine, \\ interpret, and \anold{validate} AI outcomes.\end{tabular}                  \\ \hline
\multirow{3}{*}{\anold{C3}} & \multirow{3}{*}{\begin{tabular}[c]{@{}c@{}}\anold{Workers may not be sufficiently aware} \\ \anold{about how AI can exacerbate DEI risks} \\ \anold{and perpetuate biases.}\end{tabular}}                                                    & T5          & \begin{tabular}[c]{@{}l@{}}Instruct \anold{and guide} workers \anold{through recognizing}\\ \anold{harms and benefits AI can engender on issues of}\\ diversity, equity, and inclusion in the workplace.\end{tabular} \\ \cline{3-4} 
                    &                                                                                                                                                                                                                      & T6          & \begin{tabular}[c]{@{}l@{}}\anold{Support all workers in recognizing} limitations \\ of representation in datasets and \anold{potential biases} \\ \anold{that can result.}\end{tabular}                                  \\ \hline
\multirow{7}{*}{\anold{C4}} & \multirow{7}{*}{\begin{tabular}[c]{@{}c@{}}\anold{Workers are concerned about how AI} \\ \anold{may infringe on worker rights and data privacy.}\end{tabular}}                                                                       & T7          & \begin{tabular}[c]{@{}l@{}}\anold{Educate workers on} how their worker\\ rights may be impacted \anold{by AI used in the} \\ \anold{workplace}.\end{tabular}                                                          \\ \cline{3-4} 
                    &                                                                                                                                                                                                                      & T8          & \begin{tabular}[c]{@{}l@{}}Establish straightforward grievance mechanisms \\ and train workers on how to operate these and \\ related resources.\end{tabular}                                 \\ \cline{3-4} 
                    &                                                                                                                                                                                                                      & T9          & \begin{tabular}[c]{@{}l@{}}Educate workers on the importance of,\\ and how to protect, not just workplace data \\ but their own worker data privacy.\end{tabular}                               \\ \hline
\end{tabular}
\captionsetup{width=.8\linewidth}
\caption{Mapping of \anold{AI-related challenges workers face (C1-C4)} with corresponding training \anold{topics workers desire} (T1-T9).}
\label{table_training_topics}
\end{table*}

\anold{Participants described four different challenges around workplace AI and training ideas to address these. First, they felt that some workers lack basic knowledge to recognize what AI is and how they can use it. Participants wanted training to introduce AI and expose workers to different AI tools and capabilities for work. Second, many were wary about workers' abilities to use AI properly as a tool and not a solution. Participants suggested training on related literacies needed to use AI and interpret its outcomes appropriately. Third, participants were worried that workers will overlook the potential of AI perpetuating biases or exacerbating disparities. They wanted compulsory training to raise worker awareness around DEI issues. Finally, participants were concerned about workplace AI infringing employee rights and highlighted training topics to support worker empowerment. Challenges and training topics are summarized in Table \ref{table_training_topics}. 
Below, we explain each challenge and its respective training topics by describing workers' concerns, experiences, and desires around training as it unfolded from our analysis.}


\subsection{Challenge 1: Some workers do not know what AI is or how it could help them with work tasks, which can risk them missing out on career opportunities in a rapidly evolving workforce that requires new skills.}

Perhaps because the workshop centered on discussing specific AI tools, the concern about whether workers know what AI is did not emerge. However, in interviews, several participants were skeptical if colleagues, especially older ones, know what AI is and can do. Participants \anedit{also} felt that they do not know enough about different AI tools and how to harness them for work. \anedit{A handful described AI tools they have independently sought} (e.g., Grammarly, Otter.ai, Perplexity, Revisely), and a few shared their workplaces are creating or customizing AI tools. \anedit{However, most participants described experience with only ChatGPT and/or Copilot for limited applications, i.e., text and image generation as well as translation}.\footnote{We did not ask workers to talk exclusively about generative AI. However, many of them chose to focus on generative AI tools during interviews.} 
\anold{See Table \ref{table_use_cases} for common use cases and example quotes about using AI.}

\subsubsection{Workers must start at the basics of what AI is and what it is capable of.} 
\anold{Perhaps unsurprisingly, several participants referenced going back to the "basics"} when initially answering what training workers need. While for some this equated to literacy skills to interpret AI tools correctly (\hyperref[tt3]{see Training Topic 3}), several participants pointed out that many workers \anold{and managers} need to start with what AI is, how it works, and how it can aid them \anold{(P1-3, P6, P9, P11, P13). "People are not informed. They do not have the knowledge that actually AI exists, and it can help you in certain ways" (P2).} \anold{This is also in line with prior work on what AI literacy competencies must encompass (i.e., learning how to recognize AI and its basic functions \cite{long2020ai, ng2021conceptualizing})}.

\anold{\textbf{Several attributed this need to the inherent skills gap between formal education and workplaces, especially as AI advances rapidly} \anold(P2-3, P9). P2 and P9 explained that older workers in particular may not have heard of AI, much less recognize when they are using it.} "I think a lot of us are using AI without knowing it. I didn't realize that, for example, translation is generated by AI. So, for some reason, I just thought it happens." (P9). \anold{Relatedly, a few believed some workers will need basic computer skills training before being introduced to AI (P1-2, P6, P9)}.

\anold{Thus, a training topic is necessary to tackle the fundamentals of AI and its possibilities by addressing what workers stated around (1) how workers may be wholly unaware about what AI is and (2) the importance of encouraging workers to explore it as a tool that will assist them, not replace them.}

\begin{quote}\centering
    \textit{\textbf{T1: Establish foundational understanding about what AI is}} \textit{\textbf{and what it can do \anold{to help workers.} }}
\end{quote}\label{tt1}

\subsubsection{Workers do not know enough about the range of AI tools and capabilities that are available.} 
A number of participants expressed \anold{they personally need to \textbf{learn more about what AI tools for work exist} (P1-2, P6-7, P11, P13, P15).} \anold{Participants explained this can let them ascertain \textbf{what tools are available that are helpful for their contexts.}} \anold{For example, P11 wants to use AI for tasks such as assessing future staffing needs but does not know what is available:} "I've heard, or maybe I've imagined it...if AI could help with bringing a broader perspective...that would be pretty helpful." \anold{P13 and P15 felt gaining exposure to more AI tools can also help them \textbf{identify the corresponding skills their workplaces need training around}.} P15 elaborated that AI tools themselves "influence or motivate a kind of skill."

\anold{Participants suggested workers should also have the option of follow-up training to learn in-depth about a specific product or task that interested them. For example, P5 expressed interest in how to leverage AI for creating effective messaging, and P16 suggested learning to use AI to research topics such as how other European countries have approached creating standards and policy for migrant integration.} \textbf{P14 and P16 \anold{recommended that these specialized modules be} structured for hands-on \anold{instruction.}} P16 felt the most beneficial training would begin with basic foundational training on AI followed by participants bringing their own use case to learn how to use a tool well and interpret specific inputs and outputs. 
%
%

\anold{The resulting training topic is proposed to address participants' desires for exposure to many kinds of AI tools, ideally with practicum options:}

\begin{quote}\centering
\textit{\textbf{{T2.} Expose workers to an array of emerging AI tools \anold{to learn}}} \textit{\anold{\textbf{about new tools and the skills necessary to apply them to their own work.}}}\end{quote}\label{tt2}

\anold{
\subsection{Challenge 2: Workers risk over-relying on AI and blindly trusting its outputs which can lead to harmful outcomes and loss of critical thinking skills.}
}



During the workshop, participants were wary about the extent to which AI systems can recognize situational context and support human intervention. They asked, "In what sense is the manager still at influence on the job and on AI?" and "Is it possible to interrogate the [AI's] decision made with regard to under-performance? I.e., to see the reasoning and provide human oversight?" 
In follow-up interviews, participants \anold{connected these concerns about AI systems' limitations to a worry that workers do not know how to apply critical thinking to appropriately assess AI outputs.} P3 explained that she was nearly fired at a previous workplace because her manager misinterpreted the output of an automated performance review tool. \anold{Participants also highlighted how AI systems can have limitations from lacking contexts only humans would know or have access to, and workers need to be trained on how to assess for these shortcomings.}


\subsubsection{Workers have trouble distinguishing that AI is a tool and not the answer.} 

%

\anold{\textbf{Participants were especially concerned that workers using AI will misinterpret the outcomes, become overreliant on it, or trust it blindly} (P3, P5-6, P13-14). This has been the case for a handful of participants, such as P14 who had a colleague use ChatGPT to request publications and did not check to see the document titles had been invented: "We've received another request from the field office asking us for a list of [her organization's] publications...it turned out that they were proposed by ChatGPT."
} 

\textbf{To address this, multiple participants suggested the need to review different literacy skills---\anold{numerical (P3), information (P1), data (P13-14), computer/tech (P1-2, P6, P9)}---as well as reading comprehension and critical thinking (\anold{P3, P5-6, P14})} to \anold{ensure workers can validate} the outputs of AI. P1 explained, "In my country, because we are so closed off, you would take in whatever information you receive...that's why we also have a high rate of people being scammed...we are very gullible." For P3, topics around these skills are needed so workers stop regarding AI as "the service to complete your task" and begin using it critically as an assistive tool. \anold{By using these tools regularly and thoughtlessly, P5 worried junior associates risk losing critical thinking skills.}

\begin{quote}
    "I find it quite troubling. You often hear this narrative of saying, `Oh, well, we can use AI to do some of these menial tasks and free up kind of brain power to do more interesting creative things.' But I don't really see that that's credible for very junior lawyers...they won't be doing any of those basic entry level tasks that kind of allow you to discover an organization. How it works, to build competencies. If a computer is just doing all of that for you, then I worry that they are at somewhat of a disadvantage." 

    ---P5
\end{quote}

\anold{For this training topic, we emphasize the importance of refreshing workers' literacy and critical thinking skills to ensure they are using AI in a discerning manner:}


\begin{quote}\centering
\textit{\textbf{{T3.} Refresh different literacy types and critical thinking skills}} \textit{\textbf{\anold{to orient AI as a tool to assist, not a solution to rely on.}}}
\end{quote}\label{tt3}

\subsubsection{Workers and managers must recognize limitations of AI outcomes, stemming from missing context that only humans can provide.}


\anold{Some participants made a distinction between assessing AI for objective mistakes (e.g., fake document titles) and analyzing it critically for more nuanced shortcomings.} \anold{Several described} errors \anold{that were evident to them due to their} subject matter expertise or lived experiences \anold{but would otherwise} be obscure to others \anold{(e.g., P2, P5, P9-10, P12}). For example, P2 explained that she tried using AI to brainstorm proposals for new training interventions labor officers should use. However, she realized limitations and errors in its response because it reflected information about data that she knows her country has not published online yet and addressed a problem not even applicable to her country. She reasoned this to be because "[AI] will rely on information from another country, or probably from wherever it was developed from what they envisage may be a problem that [her country] is facing, and it may not be the actual problem that we have."

While it may be simplest to ban employees from using generative AI (as P16's workplace has done), P11 shared that these are ongoing challenges her staff wants to learn how to use AI to help them with safely. She described how Workday has released an AI tool "HiredScore" to help recruiters by judging the suitability of a candidate for a role. She was \anold{hesitant to use it because} "it's not very clear to me how the system actually makes a judgment call on the suitability of the candidate" but when talking about training \anold{topics}, \anold{re-iterated her desire \textit{to} leverage these by understanding its capabilities and limitations}. \anold{These challenges that participants face inform this training topic to help workers evaluate AI outputs for contexts only knowable by humans.}

\begin{quote}\centering
\textit{\textbf{{T4.} Guide workers through \anold{using their own knowledge and expertise to}}} \textit{\textbf{\anold{critically examine, interpret, and validate} AI outcomes.}}\end{quote}\label{tt4}

\anold{
\subsection{Challenge 3: Workers may not be sufficiently aware about how AI can \anold{exacerbate DEI risks and perpetuate biases}.}
}
%
Workshop participants probed workplace AI around \textbf{fair and equitable treatment of workers.} They raised the need to evaluate disparities in AI performance across an array of criteria including different departments, pregnancy status, cases of disability, and native language spoken. A disability advocate warned how \anold{not considering these issues can result in} experiences of marginalization with technology: "Persons with disabilities, for example, are usually enthusiastic about opportunities of new technology, but can many times be very disappointed because the AI...was not designed properly. Let's say you're hard of hearing, and you have a different way of speaking and then suddenly [AI] doesn't recognize how you're speaking." 
In follow-ups, participants shared additional concerns about \anold{tools exacerbating} biases and discrimination.
\anedit{Most participants' workplaces lack formal training on AI tools, with P5 worrying: "We keep hearing about them introducing it, but nothing about how they're going to train colleagues to actually use it wisely and safely."} \anedit{Where training has been employed (P8-10, P15), it has been optional, covers general use cases of workplace AI, and focuses on how to use systems (e.g., navigating interfaces). But it has not included best practices or risks and biases to consider. P15 was one exception with his trade union developing training for members about AI and bias.}

\anold{As explained in Section \ref{method}, the workshop intentionally introduced friction around workers' rights, and several participants have experience training workers on DEI topics. Both factors may have strengthened participants' identification of risks on biases and DEI.}

\subsubsection{Workers (and managers) do not naturally consider DEI implications when using technology.}
\textbf{\anold{Some participants have observed performance discrepancies across AI tools as related to languages and sector.}} Many of our participants work in organizations that operate in at least two languages. Some participants worried about or have experienced disparities in tool performance depending on language \anold{(P5-6, P9, P12, P17)}. For instance, P5's workplace is bilingual (English and French) and serves many member states speaking different languages. She observed that tools for French-speaking users are less advanced than those for American English. This \anold{can create disparities in tool performance for employees speaking different languages} and \anold{around the quality of work they produce for member states depending on language}. P9, a training officer at an intergovernmental organization, described a unique experience alerting her that tools may be biased to certain countries or sectors. To prepare tests for interns on DEI knowledge, she asked ChatGPT, "`What are some of the advantages of diversity in the workplace?' Most of the answers that are generated is based on like corporate perspective...`it increases innovation...it increases profit.'" This led her to abandon ChatGPT as it appears to slant towards U.S.- or private-sector goals.  

\textbf{In response to these issues, several described the need for a standalone training on using AI at work and implications on DEI \anold{(P1-2, P5, P8-9, P11), with a few suggesting this should be compulsory (P5, P8-9)}.} P9 explained that when workers consider AI, "they will not really naturally think of DEI...The most efficient would be to include DEI in AI training [rather than AI topics within separate DEI training]...I think most people who would go for AI training generally [are] not necessarily interested or aware of the issues." \anold{In addition to training, two participants were interested in \textbf{tools aimed at supporting}} \textbf{DEI efforts}. P4 and P9 wanted to learn what products support workers with disabilities---"[tools that help] people...who have hearing impairment participate in our meeting and conferences." In these cases, participants shared they still need training to help them validate if these tools can deliver on DEI promises. For example, P9 was excited about and had recommended the Textio software to her colleague, P11, for its potential to use "inclusive language that attracts more women...people from...ethnic minority to apply." P11 was more cautiously optimistic, saying her team still needed to fully assess it but that they do not have a testing procedure. \anold{This leads to our suggestion for a training to support participants' desires to learn \textit{what} are the potential DEI harms/benefits that AI tools can engender and \textit{how to evaluate} products on these.}

\begin{quote}\centering
\textit{\textbf{{T5.} Instruct \anold{and guide} workers \anold{through recognizing harms and benefits AI can engender on issues of}}} \textit{\textbf{diversity, equity, and inclusion in the workplace.}}
\end{quote}\label{tt5}

\subsubsection{All workers must have a degree of awareness about how datasets can lead to biased AI outcomes.}

\anold{A number of participants raised concerns about biases and under-representation in datasets used by AI \anold{(P2, P5, P9-12, P14, P16)}.} \anold{Two took it further to stress the importance for all workers to receive some degree of training about the datasets powering AI to encourage awareness about representation biases \anold{(P14, P16)}.}
Especially if an organization's \anold{own} datasets are being used to train an AI, \textbf{workers must learn how human biases---including their own---are embedded and reinforced in AI outcomes, and \textit{how to account for/counter that}}. 

P11 reflected that past human reviewers' biases will pose risks if her organization adopts an HR screening AI based on data from their own staff: "[The software] kind of bases its judgment on what has been done in the past as well...that's the risk...if we as humans who are doing biased screening all the time, then it will replicate that."
P14 explained how even a seemingly objective task like indexing documents is subjective as it infuses the human indexer's opinion about how categories should be created. To raise workers' awareness around these non-obvious causes of bias, she recommended that employees who contribute to datasets used for training or who are creating models attend "a training about the training data, and to understand the different aspects about it...[that] the bias exists in real life." 

\anold{This reminded our team that while researchers have explored how to support} developers and data scientists in recognizing limitations of representation in training data, \textit{non-technical employees} are increasingly interacting with models through generative AI tools and should be similarly informed. Yet with convenient prompting interfaces, most may not pause to investigate details about the dataset's representation and evaluate if it has limitations or biases embedded. \anold{Though only two explicitly suggested this as a training topic, because a number of participants were worried about how a dataset's representation can lead to bias, an additional training topic we propose is:}


\begin{quote}\centering
\textit{\textbf{{T6.} \anold{Support all workers in recognizing} limitations of}} \textit{\textbf{representation in datasets and \anold{potential biases that can result.}}
}\end{quote}\label{tt6}

\anold{
\subsection{Challenge 4: Workers are concerned about how AI may infringe on worker rights and data privacy.}
}

\textbf{During the workshop, participants raised the need for workplaces to explain---even re-assess---the purpose of adopting an AI system to workers, how it works, and how it affects them.} 
Furthermore, participants were concerned about intrusions on worker data privacy \anold{from} AI systems continuously \anold{monitoring} workers. 
They also called attention to addressing workers' hesitancy around AI systems as it related to \textit{technophobia}. One participant explained not doing so risks negating the technology benefits a workplace hopes to gain: "The assumption behind it is, AI is supposed to enhance performance...but at the end of the day, if the same thing increases stress level, then it becomes counterproductive." 
In interviews, participants reflected on these challenges. Many also shared that they come from workplaces with top-down models where management selects technology for workers to use without consulting them, \anedit{such as P5 who described her management's propensity to pursue what "looks shiny and interesting" without employee input}. 

\anold{We observed that several training topics emerged in response to these challenges---privacy violations, technophobia, top-down management decisions---and relate back to supporting worker empowerment.}








\subsubsection{Workers should be proactively engaged in education around how their employee rights may be impacted by workplace AI}
\anold{Participants were most worried about how AI could negatively impact three specific worker rights:} discrimination (appraisal or wages), legality of workplace monitoring, and occupational safety and health (mental health).


\begin{itemize}
    \item \textbf{Discrimination \& wages:} \anold{Participants worried AI might discriminate against workers in performance reviews and remuneration \anold{(P1, P3-8, P10, P15)}.} P1 explained labor rights is an important topic for workers to receive training if any AI is used to measure performance---"since your performance that is calculated by AI basically affects the benefit structure"---alluding to concerns of how AI may lead to discrimination that impacts wages. \anold{Some imagined a grievance mechanism (See related \textbf{\hyperref[tt8]{Training Topic 8}}), such as P5: "You should have the opportunity to kind of view the data and to challenge it...if you think that it's not capturing the work that you're actually doing."}

    \item \textbf{Privacy \& illegitimate workplace monitoring:} \anold{Though all but P3 and P16 were uncertain if their workplace uses monitoring technology, participants categorically disapproved of such surveillance AI.} P3 and P16 were certain their workplaces do not use any due to strong data protections and work councils in their countries. On the other hand, P5 was wary her workplace was already using it without informing workers---"the capability is there if they wanted to." 
    And P7, a lawyer, shared an interesting anecdote of repeatedly reminding clients not to install cameras in workplaces for monitoring: "We always say, okay, that's not allowed...you have a good intention...but it is not the way to do it...[it's] against constitutional rights."
    
    \item \textbf{Occupational safety \& health (OSH):} Workshop participants raised concerns around AI negatively impacting workers' \anold{physical and mental} health. One participant asked whether an AI for productivity would punish workers for taking necessary breaks. Another explained that conversational agents can also present OSH concerns---"A chatbot, which can harass or discriminate or get things wrong or even just be frustrating to operate...being a mandatory thing to [use] can affect Occupational Safety and Health from a mental health perspective." \anold{In interviews, the concern around impact on OSH---related to physical health---was only mentioned by P1.}
\end{itemize}

\anold{Some specified that \textit{training} on worker rights is important (P3-9) with P3, P5, and P6 elaborating it should be required and/or recurring. The gravity of all participants' concerns leads us to suggest a specific topic to inform them of their worker rights in relation to AI.}


\begin{quote}
\centering
\textit{\textbf{{T7.} \anold{Educate workers on how their worker} rights may be impacted \anold{by AI used in the workplace.}}}
\end{quote}\label{tt7}

\subsubsection{Workers should have access to a grievance mechanism in the face of workplace AI}
HCI researchers have highlighted concerns that workplace technologies---e.g., Emotional AI (EAI)---are invasive and surveil unknowing workers \cite{roemmich2023emotion, corvite2023data}; in response, researchers have explored counter-tools to help workers combat these \cite{do2024designing}. \textbf{Yet, when asked whether workers need such training to identify when AI is being used for management or surveillance---most participants exhibited little interest.} P3 analogized this to differences in regional data privacy approaches: "How to recognize if AI is being used is \textbf{too late}, because you should have been informed...there's just a huge difference between...USA data privacy versus European." She explained that in the EU, "there has to be \textbf{\textit{need}} to collect data...You can't just...collect it because it could be useful in the future." 

\anold{In contrast to} an adversarial angle, \anold{several} participants \anold{insisted} that \anold{workplace adoption of} AI---particularly AI that impacts assessments or decisions made about employees---\textbf{must be accompanied by an understandable grievance process} \anold{(P3, P5-6, P9, P11, P15)}. 
\anold{Moreover, participants insisted} workers receive training specifically to learn how to navigate the process. "There must be a way of integrating options for employees to contest an AI-driven decision, with a very clear process...to review and also to give feedback of any decision...affecting the employee." (P15). 





\begin{quote}
\centering

\textit{\textbf{{T8.} Establish straightforward grievance mechanisms and}} \textit{\textbf{train workers on how to operate these and related resources.}
}

\end{quote}\label{tt8}
\subsubsection{Workers should not become complacent about their own data privacy.}
\textbf{Several participants shared they have mandatory training on how to protect workplace data and confidentiality using AI tools} (P7-9). This led them to consider whether workers needed training about their own data privacy. P8 reflected that surveillance technologies might be implemented for worker productivity, but the worker data collected "should not be used for over and above for what it was intended." P6 advocated for data privacy training, saying, "Workers need to show more interest about...how their personal data is treated. Most of them, as were told that is going to be used to check their attendance, [but] do not give greater importance about what happens \textit{after} their collection." \anold{P6 worried workers do not recognize that workplaces could use their data in ways they did not agree to initially.} \anold{Though only a handful of participants raised worker's personal data privacy as a separate topic to address, we felt this was still important to distinguish as a training topic given the continuing concerns around user data privacy and AI, including efforts to address data protections through regulations such as the EU's AI Act \cite{EU} and guidelines such as Peru's National Artificial Intelligence Strategy (ENIA) \cite{ENIA}.}

\begin{quote}
\centering
\textit{\textbf{{T9.} Educate workers on the importance of and how to protect}} \textit{\textbf{\anold{not just workplace data but} their own worker data privacy.}}
\end{quote}\label{tt9}

\section{Findings: Organizational and Regional Contexts Influencing Pursuit and Outcomes of Training} \label{findings_context}
As workers shared their experiences, challenges, and ideas about training for responsible AI adoption, it became clear that successful uptake and outcomes of training is highly dependent on organizational and regional contexts. Thus, even if high quality training is developed and made available for workplaces, the efficacy \anedit{will} differ based on various factors. 

\paragraph{\textbf{Change management must be grounded in empathy and trust.}}
Participants stressed that change management lacking clear communication and trust will breed a workplace of fear and/or lead to employees rejecting integration of AI. Both P8---working at a technology consulting firm---and P4---founder of an all-female staffing, training, and coaching company---shared how the phrase "digital transformation" has become a signal to workers that their organization is about to adopt new technologies and downsize. P4 explained, "what people experience in that moment are stress, sadness" because they are never explained the purpose nor "what is going to be the \textit{end} of this `transformation.'" She urged organizations to "start with empathy" because "people, who is starting with a new system...needs to see what [it] is going to be for them...not only for the company." P16 differentiated between organizational messages and direct supervisors: "Specifically if [your] \textit{\textbf{direct}} managers do not see the benefits of [AI or training] and kind of put it to the side...people might want to use [AI], but then...just continue on the way with what [they're] doing." 

Building trust also depends on whether organizations seek and meaningfully implement workers' feedback. Several participants wanted to see their organizations be more encompassing of worker perspectives. P1 pointed out the difference: "You are taught the functionality of the system and how it's going to affect your work. However...we do not have opportunities in those trainings to give feedback as to how the AI could also help us instead of just...what the organizations want...missing in these trainings is a consultation factor." On that note, P5 shared that her organization has a staff committee that is "supposed to represent the views of employees, but...there's no mechanism for really ensuring that their views are taken forward." She was disappointment to not be contacted yet for inclusion despite volunteering.


\paragraph{\textbf{Regional differences invariably produce differentials in policies, norms, and standards.}}
Participants' responses also reflected that idiosyncrasies of countries and international organizations play a role in how and whether training and responsible AI is feasible. Those from intergovernmental organizations explained that despite being headquartered in EU countries, they are not subject to domestic policies such as the AI Act, and their organizations create their own standards. Participants from \anedit{counties such as the} Netherlands exhibited more confidence than others about the privacy and protections afforded to them by their organizations. They explained they felt comfortable and aware because their countries have strong worker rights and any changes in the workplace would be championed and communicated by works councils. Participants from other countries explained they lack national policy guidance and \anedit{expressed mixed confidence} over their employers' ability in training and using AI safely.

Participants also highlighted how differences in country resource limitations play a role in their ability to access training and AI to adopt in the first place. Participants, notably from developing countries, spoke plainly about having to first address their countries' lack of internet infrastructure, slow rate of digitization, and high cost of technology. For example, P17 explained that 10\% of his country still lacks internet infrastructure.\footnote{Note: this is not equivalent to the percent of \textit{people} who lack internet access which may be higher.} P13 shared that though he was able to attend a MATLAB software training, he ultimately could not complete the assignments because the course required a paid version that he could not afford.

\section{Discussion}
\anold{In our findings, we shared} nine AI training \anold{topics} that \anold{participants} believe must be addressed to support safe adoption of AI. Below, we reflect on how these training \anold{topics} might be \anold{interpreted or} achieved---first, in consideration of the different contexts participants shared; second, by imagining how ongoing and future HCI research could be harnessed strategically; and third, in a way that is mindful of how values may become embedded in technology through training.

\subsection{Supporting Training Topics Across Different Organizational and Regional Contexts} 

\anedit{Akin to what} we shared in Section \ref{findings_context}, scholars have similarly discussed the ways organizational contexts and norms play a role in enabling responsible AI adoption \cite{liker1999perspectives, rakova2021responsible, wong2023seeing, kawakami2024situate}, including whether responsible AI resources like AI toolkits can even be adopted \cite{wong2023seeing}. Given this, we wanted to address how the feasibility of the training \anold{topics} we offer through our paper may vary based on the organizational and regional contexts our participants raised and \anedit{give} ideas to address challenges in achieving training \anold{topics}. \anold{We also share thoughts on how training topics should be interpreted for different workplace contexts.}

\subsubsection{Regional Differences in Worker Protections: Feasibility of T7-9} \label{d1_regional}
These training \anold{topics} encompass incorporating AI implications on worker rights into training (\anold{\textbf{T7}}), ensuring workers have a grievance mechanism and corresponding training (\anold{\textbf{T8}}), and educating workers \anold{about their worker data and data privacy they should be afforded} (\textbf{T9}).

We recognize that the execution of these three \anold{topics} will be directly influenced by the degree of worker protections in different regions. Specifically, those in countries with strong worker representation---e.g., Dutch works councils in the Netherlands---will have a higher likelihood of realizing worker-centered solutions such as co-designing grievance mechanisms. In contrast, this will be challenging for participants in intergovernmental organizations. We recall how these participants expressed uncertainty around how their organizations make decisions on technology to adopt and whether surveillance technologies are being used. (We did not perceive any clear trends from participants in developing countries about whether such practices would be likely or not by their workplaces.)

One idea for workers in countries without strong, formal worker representation is attending training sessions held externally to gain new knowledge. We recall how P15's trade union has begun creating its own materials on the basics of AI and AI ethics to train its members. Seeking out local trade unions, worker unions, or other external experts that offer similar topics could be one path for workers of organizations that are less keen to provide it due to lack of policies.

\subsubsection{Organizational Differences in Resources and Digitization: Feasibility of T1-3}\label{d1_organization}
These training \anold{topics} cover foundational understanding of AI (\textbf{T1}), exposure to different AI tools (\textbf{T2}), and refreshers of different literacies (\textbf{T3}).

Participants from developing countries shared how lagging technology adoption, unstable or insufficient internet infrastructure, and high cost are current barriers to entry for learning and using new AI tools. All of these pose challenges for meeting \textbf{T1} and \textbf{T2}. This should not be a burden that they bear on their own. Recall that P13 was unable to fully participate in a training because he could not afford the paid software version, but the free version had limited functions. First, training should be made affordable for participants so access to learning about these topics and tools is more open. This is already a practice in areas like academic conferences that offer sliding scales depending on the registrant's country. Organizers and facilitators should also consider how to address potential barriers to participation after participants are enrolled, such as providing full access to the software, at least for a set period to allow participation in the training and practice afterwards. We offer some related reflections in Section \ref{discussion_repo} where we discuss tools for AI training.

Additionally, a few participants were concerned about their colleagues' basic computer literacy skills, suggesting an emphasis would need to be placed on \textbf{T3}. We noticed that participants who described this need come from government or intergovernmental organizations, perhaps suggesting that there are patterns in literacy gaps or priorities across sectors, worker roles, and departments. In fact, it would benefit any organization to survey workers about the kinds of refreshers they require. This requires worker trust in their organization to answer truthfully but can support workers feeling more comfortable and better equipped before using AI. For example, perhaps numerical literacy should be emphasized for managers to ensure they properly interpret outputs, especially quantitative metrics.

\subsubsection{Interpreting Training Topics for Different Workplace Contexts}\label{d1_workplace}
When interpreting the topics we put forth, it is important that readers remember our exploration was centered on knowledge workers and the challenges and training they require around \anedit{AI, especially} general-purpose AI. This differs from past explorations on workers' onboarding needs of AI-assistant tools, typically created and designed for a specific domain, role, and/or task---e.g., medical experts making diagnoses or analyses \cite{cai2019hello, xie2020chexplain}, caseworkers reviewing child maltreatment reports \cite{kawakami2023training}.

For that reason, we caution against interpreting these training topics as a minimum set of needs for other types of workers or AI applications. Instead, what may be more appropriate is for organizations to treat these training topics as a starting point when exploring how general-purpose AI could benefit their knowledge workers, regardless of the domain---e.g., legal, medical, government. For example, knowledge workers knowing how to apply critical thinking to interpret AI outcomes \textbf{(T3)} and the importance of recognizing one's own worker data privacy \textbf{(T9)} should not change based on workplace domain (though the feasibility of pursuing it may vary depending on country or sector as we hinted in Section \ref{d1_regional}). However, training topics may see variations or additions: we imagine those in medical fields will need to be trained on implications to workplace, worker, and \textit{client} data privacy related to interacting with a general-purpose AI tool and disclosing information to it.




\subsection{A Repository of Prototyping Tools: Re-purposing Research Artefacts as AI Training Practicum Use Cases} \label{discussion_repo}



A main challenge workers face with generative AI and training is how the field of AI tools and capabilities is rapidly and constantly changing. This makes it difficult for organizations to know what to train for when management and workers may not gain exposure to new tools immediately. Several participants connected this to the daunting prospect of continuously seeking and taking training to simply stay abreast of AI tools, skills/literacies needed to use them, and related regulations and standards. 
Research on AI has likewise been fast-paced as researchers around the world explore its diverse capabilities towards real world applications \cite{zamfirescu2023johnny, park2023generative, srivastava2022beyond}. In many cases, designers and engineers are creating working prototypes and testing them in experimental settings \cite{fan2024lessonplanner, shaikh2024rehearsal, zamfirescu2023johnny, xu2024jamplate}. 
This allows them to explore AI as a design material, observing how people perceive and interact with the prototypes to learn design implications and user competencies toward productive and safe use of AI. 
We observe researchers \anold{have used several approaches when creating prototypes} including: 
\begin{enumerate}

    \item \textit{Systems to test what AI can do and how users engage with it} (\anold{e.g.,} \cite{suh2023sensecape}'s Sensecape, \cite{brade2023promptify}'s Promptify). 
    
    \item \textit{Tools to see how AI works for specific use cases} (\anold{e.g.,}  \cite{shaikh2024rehearsal}'s Rehearsal, \cite{xu2024jamplate}'s Jamplate, \cite{fan2024lessonplanner} LessonPlanner)
    \item \textit{Interfaces to support user's understanding about how AI works, related to explainability} (\anold{e.g.,} 
 \cite{jiang2023graphologue}'s Graphologue)
    
\end{enumerate}

%

\anold{The above is not intended to be a comprehensive list of approaches. However, we observe initial connections between \anedit{these} and our training topics. This gives us the idea that prototypes could be collected into a repository for organizations to leverage when training workers, emulating a hands-on approach as desired by several participants (e.g., P2, P13-16). We elaborate on each type below.} 

\anold{First, \textbf{(1) \textit{Systems to test what AI can do and how users engage with it}} would include prototypes that allow users to practice general tasks and begin learning what AI is capable of and how it can benefit them. This aligns with \textbf{T1}. This would also help participants such as P1, P2, and P14 who wanted to show management different AI tools and capabilities to encourage investment in products and related training. For instance, \citet{suh2023sensecape}'s Sensescape is a system where users practice on a generic task---i.e., trip planning---to learn how to search for and organize information; and 
\citet{brade2023promptify}'s Promptify helps users learn to create and refine prompts to generate images. While systems in this category would likely be emulating the functions of existing commercial products, the advantages of accessing them in this repository would be 1) (hopefully) a more financially feasible option than subscribing to a paid service such as OpenAI, 2) research prototypes may offer additional features to commercial products as developers test better/different ways for supporting users, and 3) some prototypes may have inherent scaffolding built-in to support users' engagement and exploration, which would be advantageous for worker training.

Next, \textbf{(2) \textit{Tools to see how AI works for specific use cases}} includes those that researchers are creating to explore specific applications for AI such as lesson planning \cite{fan2024lessonplanner}, conflict resolution \cite{shaikh2024rehearsal}, and improving design skills \cite{xu2024jamplate}. This differs from the first category because the situated use case helps users imagine AI for specialized purposes, e.g., \cite{fan2024lessonplanner}'s LessonPlanner is a tool to help teachers create lesson plans using AI. Tools in this category can supplement \textbf{T2} which aims to expose workers to an array of AI tools and features so they can seek follow-up training on specific tasks or products. \anold{In addition to prototypes, the repository could collect different tasks and examples workers want to use AI for, supporting a bottom-up approach to AI training. Then, workers can choose prototypes and use cases most aligned with their work contexts. This complements how our participants expressed specific tasks and use cases they wanted AI training on (e.g., creating effective communication).} This is also similar to recent work \anold{reporting} the benefits of workers using case studies to learn about responsible AI \cite{madaio2024learning, stoyanovich2024using, kelley2023advancing} \anold{and \citet{cai2019hello}'s suggestion for a library of domain-specific test cases to aid medical experts' onboarding of task-specific AI assistants}. \anold{A distinction with \citet{cai2019hello}'s suggestion is the worker type and AI application: here, we suggest collecting use cases for knowledge workers to use when training on general-purpose AI rather than task-specific AI.} 

Finally, \textbf{(3) \textit{Interfaces to support user's understanding about how AI works, related to explainability}} can be useful tools to accompany training on \textbf{T3} and \textbf{T4}---applying critical thinking skills and recognizing gaps from an AI's limited context. AI prototypes here can help users probe and understand the decision-making criteria and limitations of AI. This can help workers develop strategies for assessing future AI tools they come across, \anold{as desired by participants like P11 who wanted to critically assess if AI tools for employee assessment can be used responsibly.}} \anold{An example for this category is \citet{jiang2023graphologue}'s Graphologue which lets users visualize prompts and LLM responses as process flow diagrams, improving their ability to probe the logic and sources behind an LLM response.}

\anold{We believe this approach} of HCI prototypes as adaptive learning tools \anold{can respond} to challenges of ensuring a responsive workforce when AI is fast-changing, and the corresponding skills workers need can be unpredictable \cite{brynjolfsson2024navigating}. Such a repository could be executed as a standalone database or a collection integrated into a pre-existing resource like the OECD's "Catalogue of Tools \& Metrics for Trustworthy AI"\footnote{https://oecd.ai/en/catalogue/tools}. This catalogue compiles toolkits, guidelines, and measurements to help practitioners search by specific criteria for resources to create safe and responsible AI. \anold{To use this repository, we imagine organizations creating or seeking training would filter by training topic to surface relevant prototypes they can use. 

One immediate drawback would be limited tools in the repository at first.}
\anold{Additionally, it must be made clear to organizations that this is a} collection of tools not intended for deployment but for workers to practice and train on AI applications, \anold{given that prototypes are generally drafts that developers and designers use to refine ideas}.
\anold{Furthermore, a critical} practical limitation is the cost of making tools available---for example, if prototypes use commercial generative AI APIs, who will shoulder the costs? We urge researchers to consider this and related questions, but to also reflect on how a repository of publicly accessible prototypes can be a first step in addressing 1) concerns of participants from developing countries about financial burdens commercial tools present and 2) the call that research funding agencies often make to ensure funded research is accessible to the public. \anold{We offer one idea at the end of Section \ref{discussion_values} for addressing cost that researchers could consider.}

\subsection{Reflecting on Values Embedded in Technology}
\label{discussion_values}

As participants discussed training \anold{topics}, we heard from many about adapting their normal practices to use generative AI tools---for example, using English to interact with ChatGPT because their native language is not supported well or at all. 
This led us to wonder whether, what, and whose values may be reified in technology as workplaces establish norms of using and training on AI. For example, it was concerning, but unsurprising, to hear the discrepancies of how well technologies work in English compared to say, Nepali, the native language of one of the participants. It was also concerning, but unsurprising, to hear that recommendations returned from ChatGPT reflected goals of private companies, as opposed to public sectors, agencies, or nonprofits. To that end, we reflect on how the training \anold{topics} we suggested could be approached to mitigate harmful values embedded into technology due to training feedback loops by expanding on one below.

\textbf{Addressing T2: Expose workers to an array of emerging AI tools for work.} We stress that the aim of exposing workers to different tools is not to push novelty for the sake of novelty. Thus, for this training \anold{topic}, we see an opportunity for the training entity (participants suggested variations of vendors, workplace representatives, and external experts) to engage with DEI officers/subject matter experts to identify tools workers should be exposed to through training. These staff members may not know specific tools that exist, but they could make suggestions---e.g., the need to make meetings more accessible for employees with disabilities or gaps in how current tools work across different groups---that technical experts can seek out to include in training. 
This could then guide the training session with an inclusive and equitable lens to encourage workers to think more critically about what AI can do. This should be balanced with introducing workers to tools that address pressing challenges they are facing. Including such a lens can nudge the company and its workers to seek technologies that support DEI values. 

\anold{The experiences of non-native English-speaking participants also motivates us to conceptualize how to advance research for tools that \textit{do} work well in their own languages. Participants' concerns about poor performance of NLP tools on non-English languages reflect an active topic in computational linguistics research: improvement of AI-based language technologies for "under-resourced" \cite{bella2024tackling} or "lower resource" languages \cite{lignos2022toward}, i.e., languages with limited amounts of training data \cite{koehn2017six, ranathunga2023neural}. 
Furthermore, both \citet{lignos2022toward} and \citet{bella2024tackling} assert the importance of collaborating directly with language speakers for representative, ethical collection and annotations of these datasets.} \anold{One idea to address both parties' needs builds on the repository introduced in \ref{discussion_repo}. The repository could facilitate an equitable exchange of services between organizations and researchers---e.g., researchers could offer free training and use of prototypes in exchange for workers agreeing to share their data generated during training (e.g., prompts typed in Nepali or indigenous languages). Crucially, we caution against this devolving into an exploitative crowdsourcing marketplace. We point to Amara\footnote{https://amara.org/}, a volunteer-based nonprofit providing translation and captioning services, as a model to emulate instead for its people-first approach as described by \citet{gray2019ghost}.} 


\section{Limitations}
Our study has several limitations. As a qualitative study, our findings should be investigated through other methods, such as a survey study with a large number of participants. Additionally, our findings are about a set of knowledge workers and training challenges/topics to support them using AI in the workplace. 
Our participants represent a specific subset of work domains and workplaces \anold{as well. As such} there may be nuances across other \anold{worker types}, work domains, \anold{and organizations} for additional AI training topics. \anedit{We encourage future research to explore these domains, even considering how training challenges and topics may differ for knowledge workers depending on the type of knowledge work they perform}. Our interviews were held in English, but for several of our participants, English is not their native language, potentially limiting what they were able to share with us during the workshop and interviews. We also acknowledge how our scenarios presented during the workshop activity may have primed participants to think about topics that they otherwise may not have raised on their own, such as diversity, equity, and inclusion; and occupational safety and health.

\section{Conclusion}
In a workshop with 39 workers from 26 countries specializing in HR, labor law, standards creation, and worker training, we explored ideas and challenges about safely adopting AI at work. We held 17 follow-up interviews to further investigate what skills and training \anold{knowledge} workers need to achieve safe and effective AI in practice. From these, we identify nine training \anold{topics} around AI that participants expressed must be met to support them in responsible AI use. \anold{\textbf{T1} and \textbf{T2} address understanding what AI is and gaining exposure to different tools}. \anold{\textbf{T3} and \textbf{T4} address ensuring workers do not over-rely on and blindly trust AI.} \anold{\textbf{T5} and \textbf{T6} aim to make sure workers are made aware about AI implications on DEI risks and dataset biases}. \textbf{T7-T9} emphasize the need for grievance mechanisms and training on worker rights and data privacy. We reflect on how these nine training \anold{topics} can be addressed when considering the following: 1) how different contexts might impact the extent to which training will be supported by an organization, 2) offering one idea for how ongoing and future HCI research could support training, and 3) how to ensure training approaches do not embed harmful values into technology.

\begin{acks}
\anedit{We are grateful to our participants for sharing their experiences and thoughtful reflections, the facilitators we partnered with for the ITCILO workshop, and our anonymous reviewers whose careful feedback helped us improve our manuscript. Thank you to Oterhime Eyekpegha who \anedit{created the graphics used in the workshop activity}. This research was partially supported by the following: the National Science Foundation CNS-1952085, IIS-1939606, DGE-2125858, and DASS-2217721 grants; the Swedish Research Council; Good Systems, a UT Austin Grand Challenge for developing responsible AI technologies\footnote{https://goodsystems.utexas.edu}; and UT Austin’s School of Information.} 
\end{acks}

\bibliographystyle{ACM-Reference-Format}
\bibliography{references}
\clearpage
\onecolumn
\appendix
\section{Follow-Up Interview Survey Responses}

\begin{table*}[htp]
\begin{tabular}{lll}
\textbf{What is your experience with programming?} & \textbf{Responses (\#)} & \textbf{Responses (\%)} \\
No experience at all.                                     & 9                         &            52.9\%             \\
A little experience---I know basic concepts in programming.                                       &            4           &      23.5\%                     \\
Some experience---I have coded a few programs before.                                                      &           1              &            6\%             \\
A lot of experience---I code programs frequently.                                                     &             0            &               0\%          \\
No response.                                              &                 3        &                 17.6\%       
\end{tabular}
\caption{Responses to the survey question about experience with programming.}
\label{table_experience_programming}
\end{table*}

\begin{table*}[htp]
\begin{tabular}{lll}
\textbf{What is your experience with computational algorithms?} & \textbf{Responses (\#)} & \textbf{Responses (\%)} \\
No experience at all.                                     &               8           &            47\%             \\
A little experience---I know basic concepts in algorithms.                                      &        4               &     23.5\%                     \\
Some experience---I have used algorithms before.                                                     &      2                   &                  11.8\%       \\
A lot of experience---I apply algorithms frequently to my work or I create algorithms.                                                     &           0              &               0\%          \\
No response.                                              &                 3        &                 17.7\%       
\end{tabular}
\caption{Responses to the survey question about experience with computational algorithms.}
\label{table_experience_computational}

\end{table*}

\begin{table*}[htp]
\begin{tabular}{lll}
\textbf{What is your experience with workplace AI tools?} & \textbf{Responses (\#)} & \textbf{Responses (\%)} \\
No experience---I do not know if I have used AI tools at work before.                                    &        0                 &           0\%              \\
A little experience---I know examples of AI tools for work.                                       &           7              &               41.2\%          \\
Some experience---I have used AI tools at work before.                                                     &               6          &          35.3\%               \\
A lot of experience---I use AI tools at work frequently or I create workplace AI tools.                                                    &             1            &               5.9\%          \\
No response.                                              &             3            &        17.6\%                
\end{tabular}
\caption{Responses to the survey question about experience using AI tools at work.}
\label{table_experience_ai_tools}
\end{table*}

\begin{table*}[htp]
\begin{tabular}{lcc}
\textbf{How do you currently learn how to use AI tools?} & \textbf{Responses (\#)} & \textbf{Optional text entry:} \\
Chatbots (e.g., ChatGPT)                                     &              8           &                         \\
Formal Training Courses                                       &           2              &    \textit{IBM, University workshops}                    \\
Search engines (e.g., Google)                                                     &            9             &                         \\
Self-experimentation                                                    &          8               &                         \\
Video Tutorials                                             &             5            &        \textit{YouTube}                \\
Other                                             &             1            &            \begin{tabular}[c]{@{}c@{}}\textit{Office resources}\\\textit{(guidance on intranet, info sessions)  } \end{tabular}        \\
No response & 4 & 
\end{tabular}
\captionsetup{width=.8\linewidth}

\caption{Responses to the survey question about how they learn how to use AI tools. \\ \normalfont Participants could select multiple options and enter specific resources for "Formal Training Courses", "Video Tutorials", and "Other".}
\label{table_learning_resources}
\end{table*}

\clearpage

\section{Workshop Scenarios}
\begin{table*}[htp]
\begin{tabular}{|c|l|l|l|}
\hline
\textbf{Group}        & \textbf{AI Tool}        & \textbf{Issue}      & \textbf{Description}                                                                                                                                                                                                                                               \\ \hline
Example&Candidate Screening     & Bias/Discrimination & \begin{tabular}[c]{@{}l@{}}An AI hiring tool is being used by a workplace\\ to assist overworked HR staff. It extracts\\ "important" and "unbiased" information\\ about applicants from their resumes, \\ cover letters, and letters of reference.\end{tabular} \\ \hline
1&Tracking and Scheduling & Fair Pay            & \begin{tabular}[c]{@{}l@{}}A work productivity software at a workplace\\ is now automatically tracking working hours \\ and calculating corresponding wages to determine \\ worker pay.\end{tabular}                                                            \\ \hline
2&Warehouse Automation    & Unclear Termination & \begin{tabular}[c]{@{}l@{}}Warehouse workers are being asked to use \\ handheld scanners which can assign tasks, \\ track their work, and even terminate workers.\end{tabular}                                                                                  \\ \hline
3&Performance Evaluation  & Bias/Discrimination & \begin{tabular}[c]{@{}l@{}}A workplace has acquired an AI tool to \\ help managers, that uses worker data to create \\ performance reviews and custom training/upskilling \\ plans for individual workers.\end{tabular}                                         \\ \hline
\end{tabular}
\caption{Workshop Scenarios. \normalfont \\The column "Group" is the breakout group assigned for the scenario.\\ "AI Tool" refers to the technology that was depicted in the scenario. \\"Issue" is the related worker rights issue that was expressed in the scenario. \\"Description" summarizes the scenario that was presented in the session.}
\label{table_workshop_scenarios}
\end{table*}
\clearpage
\section{How Participants/Colleagues Are Using AI at Work}
\begin{table*}[h]
\begin{tabular}{|l|l|l|}
\hline
\textbf{Use Case}                                                                                                                                       & \textbf{Participants}                                                     & \textbf{Example}                                                                                                                                                                                                                                                                                                                                                        \\ \hline
\textit{Language translation}                                                                                                                           & \begin{tabular}[c]{@{}l@{}}P6, P7, P9, P10, \\ P12, P16, P17\end{tabular} &  
\begin{tabular}[c]{@{}l@{}}"The translation, I've been okay with it, because \\I'm familiar with the limitation. More frustrated\\ when I'm doing Japanese, as I mentioned to you,\\ than than French. But overall, I I wasn't expecting \\that much, so I'm quite satisfied." (P9)\end{tabular}                                                                                                                                                                                                                                                                                                                                                                        \\ \hline
\begin{tabular}[c]{@{}l@{}}\textit{Drafting text}\\ \textit{(e.g., concept notes, policies,} \\ \textit{arguments, training proposals,} \\ \textit{job descriptions)}\end{tabular} & P2, P3, P11, P13                                                                       & \begin{tabular}[c]{@{}l@{}}"An officer will go into ChatGPT and say, `Write a concept\\ note on this but that concept note needs to rely on up to \\date data of the labor officers that you have currently\\ in the country.'" (P2)\end{tabular}                                                                                                                                                                                                                                                                                                                                                                        \\ \hline
\textit{Image generation}                                                                                                                               & P4, P9                                                                    & \begin{tabular}[c]{@{}l@{}}"If we are having an event on one topic, we would \\ ask colleagues who are working on graphic design \\ to come up with an image. And they often use, \\ I think, AI for that image...`okay, I would like a \\ picture showing diverse people from different \\ backgrounds getting together for a particular topic.'" (P9)\end{tabular} \\ \hline
\textit{Brainstorming training material}                                                                                                                & P9-P12                                                                    & \begin{tabular}[c]{@{}l@{}}"{[}I have{]} given {[}it{]} learning materials that that's \\ actually used in in a training workshop. And then \\ I asked AI to...come up with multiple, multiple \\ choices test to assess the learning of this, of this\\  material." (P10)\end{tabular}                                                                                 \\ \hline
\begin{tabular}[c]{@{}l@{}}\textit{Enhancing, editing,} \\ \textit{reviewing human-draft text}\end{tabular}                                                      & P1, P4, P9                                                                & \begin{tabular}[c]{@{}l@{}}"I would type up the closing [of a submission in argument], \\and then I would ask ChatGPT to, like, enhance it." (P1)\end{tabular}                                                                                                                                                                                                                                       \\ \hline
\begin{tabular}[c]{@{}l@{}}\textit{Transcription, creating} \\ \textit{meeting agendas or notes}\end{tabular}                                                    & \begin{tabular}[c]{@{}l@{}}P1, P5, P12, \\ P17\end{tabular}               &   \begin{tabular}[c]{@{}l@{}}For example, I want to make a meeting next week, \\and we can put this---the agenda for the\\ AI...ask the ChatGPT in this case. (P1)\end{tabular}                                                                                                                                                                                                                                                                                                                                                                         \\ \hline
\textit{Researching information}                                                                                                                        & \begin{tabular}[c]{@{}l@{}}P1, P12, P13,\\ P14, P17\end{tabular}          & \begin{tabular}[c]{@{}l@{}}"Training need analysis...what kind of specified jobs \\ {[}do{]} they need in Japan side...Japan {[}has{]} 14 sectors...\\ industry like Marine and Business...so we try \\ to organize this training analysis need {[}for{]} all \\ the guidelines..." (P12)\end{tabular}                                                                 \\ \hline
\end{tabular}
\captionsetup{width=.8\linewidth}

\caption{\normalfont This table summarizes use cases participants shared about how they or colleagues are using AI in their organization, the participants who shared related use cases, and a quote to help illustrate how they are using it.}
\label{table_use_cases}
\end{table*}



\end{document}